\documentclass[prd,aps,preprintnumbers,nofootinbib]{revtex4}
\usepackage{epsfig}
\usepackage{amsmath}
\usepackage{hyperref}

\def\as{\ensuremath{\alpha_{s}}}

\def\vep{\varepsilon}

\def \ep{\varepsilon}

\def\xij{x_{ij}}

\def\vij{v_{ij}}

\def\bea {\begin{eqnarray}}
\def\eea {\end{eqnarray}}

\def\be {\begin{equation}}
\def\ee {\end{equation}}

\begin{document}

\preprint{YITP-SB-10-11}

\renewcommand{\thefigure}{\arabic{figure}}

\title{Computation of the Soft Anomalous Dimension Matrix in Coordinate Space}

\author{Alexander Mitov, George Sterman, Ilmo Sung}
\affiliation{C.N.\ Yang Institute for Theoretical Physics, Stony
Brook University, Stony Brook, New York 11794--3840, USA}

\date{\today}

\begin{abstract}
We 
complete the coordinate space calculation of the three-parton correlation in
the two-loop massive soft anomalous dimension matrix.   The full answer
agrees with the result found previously by a different approach.
The coordinate space treatment of renormalized two-loop gluon exchange diagrams 
exhibits their color symmetries in a transparent fashion.
We compare coordinate space
calculations of the soft anomalous dimension matrix with massive and massless eikonal lines
and examine its nonuniform limit  at absolute threshold.
\end{abstract}

\maketitle


\section{Introduction}

As a result of recent investigations, our knowledge of
the infrared (IR) singularities of massive gauge amplitudes has
progressed from one \cite{Kidonakis:1997gm,Catani:2000ef} to two loops
\cite{Kidonakis:2009ev,Mitov:2009sv,Becher:2009kw,Ferroglia:2009ep,Beneke:2009rj,Czakon:2009zw}.
We are now in a position to predict the single IR poles of any
two-loop amplitude with arbitrary numbers of partons  of arbitrary
masses. This helps determine the behavior of amplitudes close to
kinematic thresholds, and we now have available the two-loop
input necessary  for threshold resummation to next-to-next-to leading logarithm
for heavy quark production
\cite{Beneke:2009rj,Czakon:2009zw,Ahrens:2009uz}. Much of the new information is
encoded in the two-loop anomalous dimension matrices
for coupled massive and/or massless partons, which we refer to collectively as
the ``massive soft anomalous dimension matrix"
\cite{Kidonakis:2009ev,Mitov:2009sv,Becher:2009kw,Ferroglia:2009ep,Beneke:2009rj,Czakon:2009zw}.

A goal of the present paper is to clarify the relationship between
two calculations of a key component in the two-loop massive anomalous
dimension matrix, the color antisymmetric three-parton correlation.
Implicit expressions were presented for these correlations in
\cite{Mitov:2009sv}, partly in terms of integrals in Euclidean
space.   Subsequently an elegant  analytic expression was
presented in Ref.~\cite{Ferroglia:2009ep}. In
Ref.~\cite{Ferroglia:2009ep} calculations were carried out in
momentum space, and Ref.~\cite{Mitov:2009sv} in position space. Of
course, these two approaches should give equivalent results, and we
show below that when all contributions are taken in account, they
indeed do.   Part of our motivation in
presenting the details leading to this expected outcome is that an
apparent discrepancy in the calculations was raised in
\cite{Ferroglia:2009ep}.
This involves the diagrams called ``double exchange" in Ref.\ \cite{Mitov:2009sv}
and ``planar" in \cite{Ferroglia:2009ep}, illustrated here in Fig.\ \ref{fig:cancelpair}.
As we shall see, this apparent disagreement
arises simply because Ref.~\cite{Mitov:2009sv} exhibited results 
for these diagrams before renormalization. 
We present the remaining analysis here not as new
results, but to confirm the equivalence of the two calculations, and
because results of Ref.~\cite{Mitov:2009sv} were used in deriving
the expression for the total cross section given in
\cite{Czakon:2009zw}.  Of particular relevance was the conclusion
that the two-loop massive anomalous dimension matrix is diagonal in
the $s$-channel singlet-octet basis for pair production from
incoming light quarks or gluons at ninety degrees in the partonic
center of mass.   

We will show in Sec.\ \ref{sec:double_exchange} that apparent
discrepancies between results presented in Refs.~\cite{Mitov:2009sv}
and \cite{Ferroglia:2009ep} are entirely due to two-loop diagrams
with one-loop counterterms. These diagrams, which were not presented
in  \cite{Mitov:2009sv}, do not affect the color diagonalization  of
the anomalous dimension matrix at ninety degrees. This
diagonalization was indeed confirmed in \cite{Ferroglia:2009ep}.
It is of interest to see the simplicity of the coordinate 
space analysis, which may have applications at higher orders
\cite{Dixon:2008gr,Becher:2009cu,Gardi:2009qi}.

In Sec.\ \ref{sec:massless}, we discuss similarities and differences
encountered in the coordinate space calculations 
when all lines are massless. Here, we present a discussion in which
both one- and two-loop massless integrals are regularized dimensionally, and
reproduce from a purely coordinate space analysis 
the absence of three-eikonal color correlations, a result found in \cite{Aybat:2006mz}, 
which relied on arguments in both momentum and coordinate space.

Finally, in Sec.\ \ref{sec:threshold}, we discuss the
limit of absolute threshold (pair creation at rest).
Reference \cite{Ferroglia:2009ep} discovered a non-uniform limit of the anomalous dimension matrix at
absolute threshold, when expressed in terms of the scattering angle.
Here we rederive the result found
Ref.~\cite{Ferroglia:2009ep}, using the methods introduced in
Ref.~\cite{Mitov:2009sv}.  We show how the
singularity structure of the integrals
restricts the non-uniform limit to a region of momentum space where the eikonal
approximation fails.    This 
limit of the soft anomalous dimension matrix
is not directly relevant to the
total pair production cross section  \cite{Beneke:2009rj,Czakon:2009zw}.
For completeness, explicit expressions for the relevant
integrals are exhibited in an appendix.

\section{Two-loop Double Exchange Diagrams}
\label{sec:double_exchange}

The formalism for the computation of soft anomalous dimension
matrices has been discussed extensively in
Refs.~\cite{Kidonakis:1997gm} and elsewhere.
The relevant diagrams consist
of radiative corrections to  incoming and outgoing  partonic
propagators in eikonal approximation. The resulting diagrammatic
expressions are scaleless, and vanish in dimensional regularization.
Their ultraviolet poles, however, define counterterms and
hence the anomalous dimensions we are after
\cite{Kidonakis:1997gm,Aybat:2006mz}. 
In this formalism, we factorize the full eikonal amplitude into a
soft function, with one logarithmic singularity
per loop and all nontrivial color exchange, and color-diagonal
jet functions.   This factorization ensures that in the limit of vanishing
eikonal masses the soft function is free of collinear singularities \cite{Sterman:2002qn,Aybat:2006mz}.
A natural generalization of this procedure
to eikonals of arbitrary mass is the scheme employed in Ref.\ \cite{Czakon:2009zw},
in which the jet  functions are defined as the square roots of the low-mass limits of the corresponding
form factors \cite{Mitov:2006xs,Gluza:2009yy}.\footnote{We note that Ref.\ \cite{Ferroglia:2009ep} analyzes the massive 
amplitude directly, without subtractions, but direct comparison can
still be made, because by construction subtractions do not
affect color exchange.}

\begin{figure}
{\epsfxsize=4 cm \epsffile{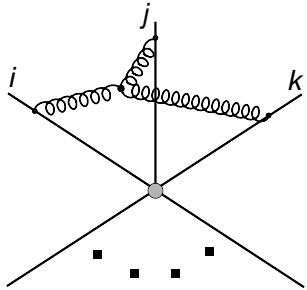}\\
 \caption{Non-planar diagram for the two-loop
 anomalous dimension matrix.}  
 \label{nonplanar}}
\end{figure}

The most challenging two-loop contribution to the massive anomalous
dimension matrix is the set of non-planar diagrams with a three-gluon
vertex connecting three  eikonals, of the type shown in
Fig.~\ref{nonplanar}. Although these diagrams are rather
complex, they have no UV-divergent subdiagrams.    It was shown in
Ref.~\cite{Aybat:2006mz} by a momentum-space change of variables, that
the ultraviolet poles of these diagrams vanish
when all three of the interacting eikonals
are massless.   This result was rederived in Ref.\ \cite{Mitov:2009sv}
in coordinate space and extended to diagrams with  two massless interacting eikonals,
while a manifestly nonvanishing 
integral representation was given for the UV pole 
when all three interacting eikonals are massive.
An analytic expression for the ``non-planar" diagrams like Fig.~\ref{nonplanar}
with massive external partons was derived in
Ref.~\cite{Ferroglia:2009ep}. Surprisingly, it turned out to
be very simple (see appendix). In numerical tests, it
agrees with the integral representation given in
Ref.~\cite{Mitov:2009sv}.
\begin{figure}
{\epsfxsize=4 cm \epsffile{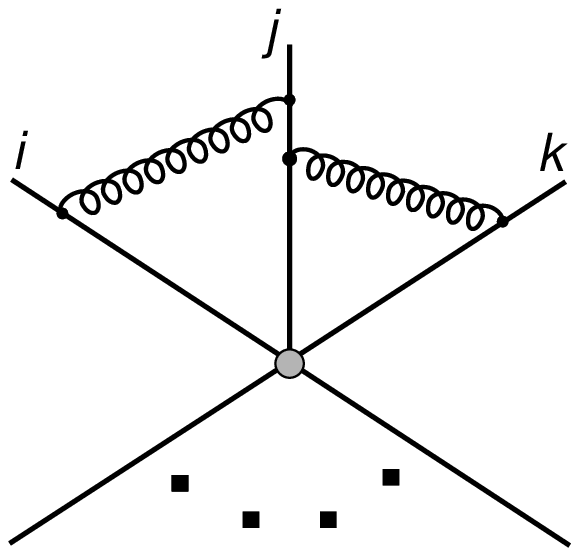} \quad \quad
\epsfxsize=4 cm \epsffile{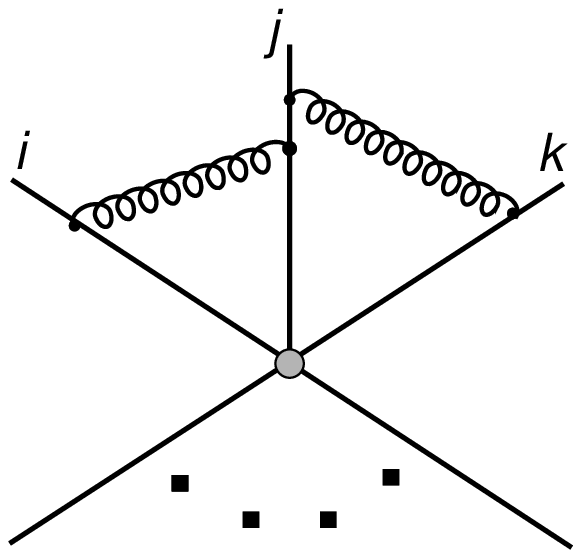} \\
\hbox{ \hskip 6.3 cm (a) \hskip 4.1  cm \quad (b) } \caption{Double
exchange diagrams discussed in the text. The
shaded circle represents an $n$-eikonal  vertex.}  
\label{fig:cancelpair}}
\end{figure}

In this section, we revisit the double-exchange diagrams of 
Fig.\ \ref{fig:cancelpair} in coordinate space, including their renormalization. 
As explained in
Ref.~\cite{Mitov:2009sv} and confirmed below, the contributions from the two-loop
double exchange diagrams are fully symmetric in color
after combining pairs like those shown in
Fig.~\ref{fig:cancelpair}.
They contain, however, subdiagrams with UV poles, and thus require one-loop
counterterms.   These counterterms were not considered explicitly in
Ref.~\cite{Mitov:2009sv}. 
\begin{figure}
{\epsfxsize=4 cm \epsffile{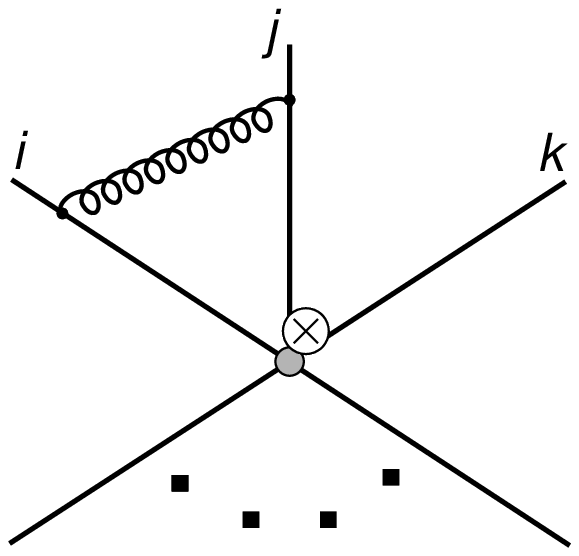} \quad \quad
\epsfxsize=4 cm \epsffile{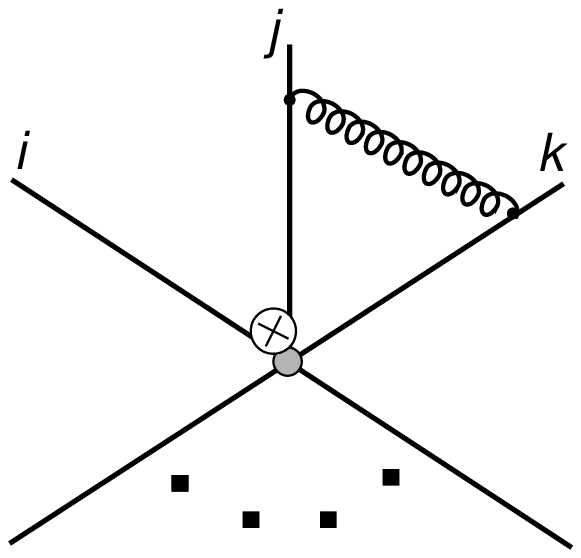} \\
\hbox{ \hskip 6.3 cm (a) \hskip 4.1  cm \quad (b) }
\caption{Order $g^4$ diagrams containing one-loop UV counterterm.}
\label{fig:UVct}}
\end{figure}
In Fig.~\ref{fig:UVct} we show the renormalization of the double exchange
diagrams of Fig.\ \ref{fig:cancelpair}, with divergent 
loops replaced by counterterms. In Fig.~\ref{fig:UVct}a, for example, the
counterterm depends on the pair of momenta $p_k,p_j$ and the
remaining one-loop diagram  on $p_i$ and $p_j$.
\begin{figure}
{\epsfxsize=4.3 cm \epsffile{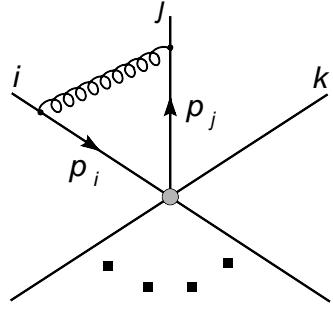}\\
 \caption{A gluon exchanged between an incoming quark $i$ and an outgoing quark $j$ that contributes to Eq.~(\ref{eq:D-def}).} 
  \label{oneloop}}
\end{figure}

To determine the counterterms in our coordinate
space formalism, and to establish notation, we review the one-loop case,
illustrated in Fig.\ \ref{oneloop}.   The counterterms, in
turn, are found from the poles of one-loop diagrams. Following the
notation of Ref.~\cite{Mitov:2009sv} (see also Appendix B of
Ref.~\cite{Aybat:2006mz} for the same calculation in slightly
different notation), we introduce the scalar propagator in
coordinate space,
\bea \Delta(x-y) &\equiv& i\int d^{4-2\vep}k\ e^{-ik\cdot (x-y)}\; \frac{1}{k^2+i\epsilon}
\nonumber\\
 &=& -~\frac{\Gamma(1-\vep)}{4\pi^{2-\vep}}
\frac{1}{\left( (x-y)^2-i\epsilon \right)^{1-\vep}} 
\nonumber\\
&\equiv& - \ c(\vep)\,
\frac{1}{4\pi^2}\frac{1}{\left( (x-y)^2 - i\epsilon\right)^{1-\vep}} 
\, ,
\label{Deltadef} \eea
where the  third relation defines the constant $c(\vep)=1+{\cal
O}(\ep)$. 

The coefficient of $\alpha_s/\pi$ in the
one-loop correction corresponding to Fig.\ \ref{oneloop} is a matrix in color space,
\footnote{This factor is referred to as the ``velocity factor" in
Ref.~\cite{Aybat:2006mz}. Because all partonic lines are eikonal, we
need not distinguish between velocities and  momenta for the
external lines.}
\begin{eqnarray}
{\cal M}^{(1)}(p_i,p_j,\ep)
&=&
\ c(\vep)\, \left ( {\bf T}_i\cdot {\bf T}_j\right)
\int_0^\infty d\lambda_j \int_0^\infty d\lambda_i { p_i\cdot p_j\over \left[
(\lambda_j p_j - \lambda_i p_i)^2-i\epsilon\right]^{1-\ep} }\, \bigg |_{UV} \nonumber\\
&=& \ c(\vep)\,  \left ( {\bf T}_i\cdot {\bf T}_j\right)\, (p_i\cdot p_j) \int_0^\infty
{d\lambda_j \over \lambda_j^{1-2\ep}} \int_0^\infty {d\sigma \over
\left[ (p_j - \sigma p_i)^2-i\epsilon\right]^{1-\ep} }\, \bigg |_{UV}
\nonumber\\
&=& \ c(\vep) \,  m_j^{2\ep}\, \left ( {\bf T}_i\cdot {\bf T}_j\right)\,
\frac{1}{2\ep}\, I^{(-1)}(p_i,p_j)
 \, .
\label{eq:D-def}
\end{eqnarray}
The color factors associated with the
gluon exchange are represented in the basis-independent notation
introduced in Ref.\ \cite{catani96} and generally employed in
analyses of soft anomalous dimensions.
The final relation in Eq.\ (\ref{eq:D-def})
shows the contribution of the UV pole to
the one-loop anomalous dimension matrix in terms of the function
\begin{eqnarray}
I(p_i,p_j,\ep) & \equiv & (p_i\cdot p_j)\, m_j^{-2\ep} \int_0^\infty
d\sigma\ \frac{1}{(m_j^2 - 2\sigma p_i\cdot
p_j+\sigma^2m_i^2 - i \epsilon)^{1-\ep}} \nonumber\\
&=& I^{(-1)}(p_i,p_j) + \ep I^{(0)}(p_i,p_j) + {\cal O}(\ep^2)\, ,
\label{Idef}
\end{eqnarray}
with $m_{i,j}^2 = p_{i,j}^2 \neq 0$.  The massless limit is explored
in Sec.\ \ref{sec:massless}.  For the purposes of this discussion,
we may take $-p_j\cdot p_i>0$, corresponding
to one line incoming and the other outgoing. In this case, $I(p_i,p_j,\ep)$ is real.
The explicit results for the
functions $I^{(-1,0)}(p_i,p_j)$ for both
signs of the invariant are given in the appendix.

It is important to note that the prefactor in Eq.\ (\ref{eq:D-def}) is
arbitrary at ${\cal O}(\varepsilon)$, because it reflects the choice we make
to define the integral over
$\lambda_j$, a variable with dimensions of length squared.
In effect, our choice for these functions constitutes a minimal subtraction scheme.
In particular, the factor $m_j^{2\vep}$ in (\ref{eq:D-def}) and following
expressions is kept as a bookkeeping device
only, because it matches the definition
of $I(p_i,p_j,\ep)$.  For the purposes of calculating 
anomalous dimensions it is only necessary that the
prefactors for all such integrals be defined in the same way.
Then, as we will see below, overall factors like
$c(\vep)$ and $m_j^{2\vep}$ will not affect our results
for the dimensionless pole terms of ${\cal M}^{(1)}$.

The one-loop soft anomalous dimension matrix is defined only up
to color-diagonal contributions associated with its eikonal lines, which
absorb collinear logarithms and are factorized from the soft function.
In any such factorization scheme, the one-loop anomalous dimension 
matrix is found from the residues of the poles $1/(-2\vep)$ in the one-loop amplitudes of the soft function
\cite{Aybat:2006mz}.
As noted above, we use the scheme of Ref.\  \cite{Czakon:2009zw}, 
in which the soft function is the full eikonal amplitude divided by the low-mass limit of the square root of the
form factor for each external eikonal line.
The one-loop soft anomalous dimension matrices found
 from the  ${\cal M}^{(1)}_{ij}$ are then given by
\bea
 {\bf \Gamma}_S^{(1)} &=& -\ \frac{1}{2} \sum_{j} \sum_{i \neq j} \left ( {\bf T}_i\cdot {\bf T}_j\right)
 \, I^{(-1)}(p_i,p_j) \ -\ \frac{1}{2}\, \sum_j C_j   \, \ln\left(\frac{\mu^2}{m_j^2}\right)
 \nonumber\\
 &=& -\ \frac{1}{2} \sum_{j} \sum_{i \neq j} \left ( {\bf T}_i\cdot {\bf T}_j\right)
 \, \left[\, I^{(-1)}(p_i,p_j) \ -\ \ \ln\left(\frac{\mu^2}{m_i\, m_j}\right)\, \right]\, 
 \, ,
 \label{Gamma1}
 \eea
 where in the first line we subtract for each external eikonal
 a term that is collinear-singular in the massless limit,
 and where $C_q=C_{\bar q}=C_F,\, C_g=C_A$ refer to the color
 content of the eikonal lines.   This 
 procedure eliminates all collinear logarithms,
as may be readily verified from the expressions for $I^{(-1)}$ 
given in the appendix.
To verify the second line of Eq.\ (\ref{Gamma1}), we recall the color identities, \cite{catani96}
\bea \sum_i {\bf T}_i=0\, , \quad {\bf T}_j\cdot {\bf T}_j = C_j\, ,
\label{coloridentity} 
\eea
with $i=F,A$.   We are now ready for an analysis  of the double-exchange diagrams.

The coefficients of $(\as/\pi)^2$ in the two-loop exchange
diagrams of Fig.\ \ref{fig:cancelpair}a, involving lines $i$, $j$
and $k$, can be written in terms of the coordinate space integrals, $I$, Eq.\ (\ref{Idef}), as
\bea
{\cal M}^{(2)} (p_i,p_j,p_k, \ep) &=& c^2(\ep)m_j^{4\ep}\,
\left ( {\bf T}_i\cdot {\bf T}_j\right)\left( {\bf T}_j\cdot {\bf
T}_k\right ) \int_0^\infty
\frac{d\lambda'_j}{\lambda_j'{}^{1-2\ep}}\, I(p_i,p_j,\ep)\ \int_0^{\lambda'_j}
\frac{d\lambda_j}{\lambda_j^{1-2\ep}} \,
I(p_k,p_j,\ep) \, \bigg |_{UV} \, , 
\label{M2integrals}
\eea
where the variables $\lambda_j$ and $\lambda_j'$ represent the locations
of the vertices along eikonal line $j$, to which two gluons are
attached, as in Fig.\ \ref{fig:cancelpair}. The ordered pairs of arguments
$p_i,p_j$ ($p_j,p_k$) denote the eikonal lines connected by the  outer
(inner) gluon exchange of the diagram.
The double pole of Eq.\ (\ref{M2integrals}) is associated with the independent limits $\lambda_j\rightarrow
0$ and $\lambda'_j\rightarrow 0$ of distances from the origin to
the vertices along the $j$ eikonal.   
The factors $I(p_i,p_j,\epsilon)$ and $I(p_k,p_j,\epsilon)$
are the result of integrals like Eq.\ (\ref{Idef}), which are over the ratios of
origin-to-vertex distances.   These integrals
converge at both zero and infinity for fixed masses and momentum
invariants.   Thus, the individual poles arise from independent scaling of the
two vertices of the $ij$- or $kj$-exchange in Fig.\ \ref{fig:cancelpair} to the origin.
Any remaining single pole represents a
scaling of the four vertices of the entire diagram together to the origin.   
The latter defines the diagram's
contribution to the two-loop anomalous dimension.

The UV poles of the $\lambda_j$ integrals  in Eq.\ (\ref{M2integrals}) can readily be
isolated, because the functions $I(p_i,p_j,\ep)$, defined in
Eq.~(\ref{Idef}), are independent of the $\lambda_j$'s. We find,
\bea {\cal M}^{(2)} (p_i,p_j,p_k, \ep) &=& c^2(\ep)m_j^{4\ep}\,
\left ( {\bf T}_i\cdot {\bf T}_j\right)\left( {\bf T}_j\cdot {\bf
T}_k\right )\ \left[ \frac{1}{8\ep^2}\, I^{(-1)}(p_i,p_j) \,
I^{(-1)}(p_k,p_j) \right. \nonumber\\
&\ & \left. \hspace{20mm} \ +\ \frac{1}{8\ep}\, I^{(-1)}(p_i,p_j)\,
I^{(0)}(p_k,p_j)\, \ +\   \frac{1}{8\ep}\, I^{(0)}(p_i,p_j)\,
I^{(-1)}(p_k,p_j) \right]\, , 
\eea
with both double and single poles, in a form that is clearly
symmetric in the pairs $ij$ and $kj$ \cite{Mitov:2009sv}.

The complete set of two-loop diagrams corresponding to Fig.\
\ref{fig:cancelpair} includes as well counterterms for the
UV-divergent subdiagram consisting of the (inner) exchange 
as shown in Fig.\ \ref{fig:UVct}.  For Fig.\ \ref{fig:UVct}a, this one-loop counterterm is simply $-{\cal
M}^{(1)}(p_k,p_j,\ep)$, defined by Eq.\ (\ref{eq:D-def}). For
convenience, we include the factor $c(\vep)$ as part of the
renormalization scheme.  The diagrammatic representation of the
corresponding two-loop contribution is shown in Fig.\
\ref{fig:UVct}a, and is given by
\bea {\cal M}^{(2,ctr)} (p_i, p_j,p_k,\ep) &=&
-c^2(\ep)m_j^{4\ep}\, \left ( {\bf T}_i\cdot {\bf T}_j\right)\left(
{\bf T}_j\cdot {\bf T}_k\right )\ \int_0^\infty
\frac{d\lambda'_j}{\lambda_j'{}^{1-2\ep}}\ I(p_i,p_j,\ep)\,
\frac{1}{2\ep}\, I^{(-1)}(p_k,p_j) \bigg |_{UV} \\
&\ & \hspace{-20mm}
 =
 \ -c^2(\ep)m_j^{4\ep}\, \left ( 
 {\bf T}_i\cdot {\bf T}_j\right)\left( {\bf T}_j\cdot {\bf T}_k\right )\
\left[\left(\frac{1}{2\ep}\right)^2\, I^{(-1)}(p_i,p_j) \,
I^{(-1)}(p_k,p_j)\ +\ \frac{1}{4\ep}\, I^{(0)}(p_i,p_j) \,
I^{(-1)}(p_k,p_j)\, \right]\, , \nonumber
\eea
again providing both single- and double-pole contributions.
Combining Fig.\ \ref{fig:cancelpair}a with its one-loop counterterm, Fig.\ \ref{fig:UVct}a,
we find
\bea {\cal M}^{(2)}  (p_i, p_j,p_k,\ep) + {\cal M}^{(2,ctr)} (p_i,
p_j,p_k,\ep) &&=
\ c^2(\ep)m_j^{4\ep}\, \left ( {\bf T}_i\cdot {\bf
T}_j\right)\left( {\bf T}_j\cdot {\bf T}_k\right )\
 \left( -\frac{1}{8\ep^2}\, I^{(-1)}(p_i,p_j) \,
I^{(-1)}(p_k,p_j)\ \right. 
\nonumber\\
&&
\hspace{-10mm}
\left. -\frac{1}{8\ep}\, \left[\, I^{(0)}(p_i,p_j) \,
I^{(-1)}(p_k,p_j)\, -\,I^{(-1)}(p_i,p_j) \, I^{(0)}(p_k,p_j)\,
\right] +{\cal O}(\ep^2)\right)\, . \label{eq:Mpluscounter}
\eea
In this expression, we see directly that the double pole is
symmetric in the momentum pairs $p_i,p_j$ and $p_k,p_j$, while the single pole is
purely antisymmetric.

We next combine the foregoing result with the diagram in which the
roles of the pairs $ij$ and $jk$ are exchanged, as in Figs.\
\ref{fig:cancelpair}b and \ref{fig:UVct}b. Including color factors, we
find the following pole and color structure,
\bea {\cal M}^{(2)} (p_i, p_j,p_k,\ep) + {\cal M}^{(2,ctr)}  (p_i,
p_j,p_k,\ep) + \left( i \leftrightarrow k\right) 
&=& -\left\{ {\bf
T}_i\cdot {\bf T}_j, {\bf T}_j\cdot {\bf T}_k\right\}\
\frac{c^2(\ep)m_j^{4\ep}}{2(2\ep)^2}\, I^{(-1)}(p_i,p_j) \,
I^{(-1)}(p_k,p_j)\, \nonumber\\
&\ & \hspace{-48  mm} +\ \left[ {\bf T}_i\cdot {\bf T}_j, {\bf
T}_j\cdot {\bf T}_k \right]\ \frac{c^2(\ep)m_j^{4\ep}}{8\ep}\, \left[\,
I^{(-1)}(p_i,p_j) \, I^{(0)}(p_k,p_j)\, -\,I^{(0)}(p_i,p_j) \,
I^{(-1)}(p_k,p_j)\, \right]\, . 
\label{bothM}
\eea
We recognize that the single pole terms in this expression enter entirely with antisymmetric color factors,
while the double poles are entirely symmetric in color \cite{Mitov:2009sv}.

The combination of all such three-eikonal double exchanges (DE) can
now be written down for an arbitrary number
of eikonal lines. The double pole terms of Eq.\ (\ref{bothM}) 
are cancelled by double poles of two-loop counterterms, which we see
are the expansion of exponentiated one-loop counterterms.
This leaves the
single-pole, color-antisymmetric terms. Recalling
\cite{Aybat:2006mz} that the contributions of these diagrams to the
two-loop anomalous dimension matrix is the residue of the
single pole, $-1/(4\ep)$, we find
\begin{eqnarray}
{\bf\Gamma}_S^{(3E,{\rm DE})} = \left({\as\over \pi}\right)^2~{1\over
2}~ \sum_{i>j>k=1}^n ~ if^{abc}{\bf T}_i^a{\bf T}_j^b{\bf
T}_k^c~\sum_{I,J,K=(i,j,k)} ~\ep^{IJK}~ I^{(-1)}(p_I,p_J) \,
I^{(0)}(p_K,p_J)\, , 
\label{eq:Gamma-planar-ct}
\end{eqnarray}
where we have exhibited the antisymmetric color structure explicitly
by carrying out the commutators in Eq.\ (\ref{bothM}). In this sum, the $\ep$-expansions of
the overall factors Eq.\ (\ref{bothM}), $c^2(\vep)m_j^{4\ep}$, cancel because of the antisymmetry in
(\ref{eq:Gamma-planar-ct}), which is invariant under any
modification $I^{(0)}(p_i,p_j)\rightarrow I^{(0)}(p_i,p_j)+{\rm
const}\times I^{(-1)}(p_i,p_j)$. 
In fact, we can generalize this result to any overall
rescaling of the integrals $I$,
\bea
I(p_i,p_j,\ep) \rightarrow C(p_i,p_j,\vep)\, I(p_i,p_j,\ep)\, ,
\label{eq:Iambiguity}
\eea
where $C$ has an expansion, $C(p_i,p_j,\vep)= 1+\vep C^{(1)}(p_i,p_j) + \dots$.
As can be seen immediately from Eq.\ (\ref{bothM}),
any such overall rescaling has no effect on 
the antisymmetric single pole terms of the two-loop anomalous dimension.
Such a rescaling, of course, does change the symmetric color terms
at the single-pole level,
and should be thought of as a change in scheme for the soft function.
Equation (\ref{eq:Gamma-planar-ct}) is in the form found in
Ref.\ \cite{Ferroglia:2009ep}.   
The exact expressions for the integrals $I(p_i,p_j,\ep)$, in terms
of which our results have been expressed so far, are given
explicitly in the appendix for both space- and
time-like kinematics.  These results
confirm the consistency of the two approaches in
Refs.~\cite{Mitov:2009sv} and \cite{Ferroglia:2009ep} that we set
out to establish. 

To conclude this section, we note for completeness 
the form of the full three-eikonal anomalous dimension matrix, which
is found by combining the results from the double-exchange diagrams,
Eq.~(\ref{eq:Gamma-planar-ct}), with the three-gluon diagram, Fig.\
\ref{nonplanar}.    The latter has the same color structure, 
and can be incorporated into Eq.\ (\ref{eq:Gamma-planar-ct})
by simply adding terms to the factors $I^{(0)}$ in Eq.\ (\ref{eq:Gamma-planar-ct})
 \cite{Ferroglia:2009ep},
\begin{eqnarray}
{\bf\Gamma}_S^{(3E)} = \left({\as\over \pi}\right)^2~{1\over 2}~
\sum_{i>j>k=1}^n if^{abc}{\bf T}_i^a{\bf T}_j^b{\bf T}_k^c \,
\sum_{I,J,K=(i,j,k)} ~\ep^{IJK}~ I^{(-1)}(p_I,p_J)\, \left[
I^{(0)}(p_K,p_J) + I_{3g}(p_K,p_J) \right]\, .
\label{eq:Gamma-full-ct}
\end{eqnarray}
The explicit result for $I_{3g}$ can be found in the appendix.
Finally, we note that to construct the full two-loop anomalous dimension matrix
for an $n$-point amplitude one needs to combine
Eq.~(\ref{eq:Gamma-full-ct}) with the
diagrams in which gluons connect to only two eikonals.
The latter, which have the same color structure
as the one-loop matrix,   have been given explicitly  in Ref.~\cite{Czakon:2009zw}.

\section{Massless Eikonals in Coordinate Space}
\label{sec:massless}

The diagrams treated in the previous section, of course, also appear
in the calculation of the anomalous dimension matrix with 
massless eikonal lines \cite{Korchemskaya:1992je}. In Ref.\ \cite{Aybat:2006mz}, the same
double exchange diagrams considered above, Figs.\
\ref{fig:cancelpair}a,b, were shown for the massless case to give
vanishing contributions to the two-loop anomalous dimension matrix
using a momentum-space analysis. 
That is, their only contributions are cancelled by two-loop
counterterms that are expansions of exponentiated one-loop
counterterms. It is worthwhile examining how this result is
maintained in coordinate space when taking into account the counterterm diagrams of Fig.\
\ref{fig:UVct}.

We begin by recalling that the soft function for massless lines must
be defined to eliminate collinear poles on a
loop-by-loop basis.   We choose the form employed in Ref.\
\cite{Aybat:2006mz}, and mentioned above,
where the soft function is defined as the ratio of
the full eikonal amplitude to the product of the square roots of
color-singlet form factors, one for each eikonal line.   
This is the analog of the color diagonal subtraction in Eq.\ (\ref{Gamma1}) in the massive case.
We will not need the details of this construction here, but observe that it
requires that at each loop order, double poles appear with only
color-diagonal coefficients after the sum over diagrams. We will see
this feature emerge explicitly below, as we discuss the color
structure of one- and two-loop exchange diagrams, beginning with the
massless analog of the one-loop diagram, Eq.\ (\ref{eq:D-def}).

For computations involving massless eikonals, we continue to use momenta
$p_i$ in place of dimensionless four-velocities.
Because the integrands that define the amplitudes are
scaleless in the velocities, this is a trivial substitution before
integration.    
For $p_i^2=p_j^2=0$
we then have
\begin{eqnarray}
\mu^{2\vep}\, {\cal M}^{(1)}(p_i\cdot p_j,\ep)_{p_i^2=p_j^2=0} 
&=&  
\mu^{2\vep}c(\vep)\, \left ( {\bf T}_i\cdot {\bf T}_j\right) \int_0^\infty
d\lambda_j \int_0^\infty d\lambda_i {p_i\cdot p_j\over
\left[(\lambda_j p_j - \lambda_i p_i)^2 - i \epsilon \right]^{1-\ep} }\,
\bigg |_{UV} \nonumber\\
&=&  
\mu^{2\vep}c(\vep)\, \left ( {\bf T}_i\cdot {\bf T}_j\right) \int_0^\infty
d\lambda_j \int_0^\infty d\lambda_i {p_i\cdot p_j\over
\left[-2p_i\cdot p_j \lambda_i\lambda_j  - i \epsilon \right]^{1-\ep} }\,
\bigg |_{UV}
 \, ,
\label{firstcalIdef}
\end{eqnarray}
where we have multiplied by $\mu^{2\vep}$ so that both sides of this
expression are dimensionless.   This will help motivate our choice of renormalization
scheme for this integral.
As in our discussion of the massive case, we treat $-2p_i\cdot p_j$ as a positive quantity,
and then analytically continue to negative values.   
The coordinate space expression in (\ref{firstcalIdef})
has two scaleless integrals rather than one, as in Eq.\ (\ref{eq:D-def}).
The treatment of these integrals requires some discussion,
even in the one-loop case.

The presence of mixed infrared and
collinear singularities in a Lorentz-invariant 
integral make it impossible to maintain at the same time
the correct power behavior under the scaling of momenta
and invariance under rescalings of the velocities, a feature sometimes
referred to as the ``cusp anomaly" \cite{Gardi:2009qi}.    As a result, the prescription to
identify the ultraviolet pole does not, as in the massive case,
uniquely fix the coefficient of that pole.   
We can organize 
these ambiguities by  changing variables from the
$\lambda$'s, which have dimensions  of length squared,
to dimensionless variables.    This is a unique change of
variables in this case, if we insist that the integral (\ref{firstcalIdef})
depend only on $p_i\cdot p_j$.
We are therefore led to the dimensionless variables
\bea
l_j \equiv \lambda_j\ (-2p_i\cdot p_j)\, , \quad 
l_i \equiv \lambda_i\ (-2p_i\cdot p_j)\, .
\label{elldef}
\eea
After this change of variables, the dimensional
content of the integral is manifest.    To isolate the
ultraviolet pole, we use the symmetry between
$l_i$ and $l_j$, and write
\begin{eqnarray}
\mu^{2\vep}\, {\cal M}^{(1)}(p_i\cdot p_j,\ep)_{p_i^2=p_j^2=0} 
&=&  
-\ \left ( {\bf T}_i\cdot {\bf T}_j\right) 
c(\vep)\, \left(\frac{\mu^2}{-2p_i\cdot p_j}\right)^{\vep}\; 
\int_0^\infty \frac{dl_j}{ l_j^{1-\vep}} \int_{l_j}^\infty \frac{d l_i}{l_i^{1-\vep}} 
\bigg |_{UV} 
 \, .
\label{calIdef}
\end{eqnarray}
The $l_i$ integral is carried out at fixed $l_j$ for
infrared regularization ($\vep<0$), to give a simple
pole.   Next, the ultraviolet pole of the $l_j$ integral is isolated.
This results, of course, in an expression with
both single and double poles,
\bea 
\mu^{2\vep}\,
{\cal M}^{(1)}(p_i\cdot p_j,\ep)_{p_i^2=p_j^2=0} 
&=& 
\left ( {\bf T}_i\cdot {\bf T}_j\right) 
c(\vep)\, \left(\frac{\mu^2}{-2p_i\cdot p_j}\right)^{\vep}  \frac{1}{2\vep^2}
\nonumber\\
&=&
\frac{1}{2}\, \left ( {\bf T}_i\cdot {\bf T}_j\right)\, \left(
\frac{1}{\ep^2}\ -\frac{1}{\ep} \ln\left(\frac{-2p_i\cdot p_j}{\mu^2}\right)\right) + \dots
\, .
 \label{eq:M1m=0-def}
\end{eqnarray}
Momentum-dependence appears only
in single poles, while double poles are momentum-independent.
In either form of (\ref{eq:M1m=0-def}), we can revert to
a velocity dependence by choosing $\beta_i\equiv p_i/\mu$.
Keeping the momentum-dependence, we can identify
the first line of Eq.\ (\ref{eq:M1m=0-def}) as the eikonal
contribution to the familiar function ${\bf I}^{(1)}$ defined in \cite{Catani:1998bh}.
As in the massive case above, the choice between the
first and second lines is a choice of renormalization scheme.
In fact, we could use the first line to define an alternative scheme
to define the soft function.   In this discussion,
however, we stick with the choice of pure poles, as in Ref.\ \cite{Aybat:2006mz}.

As in the example of Eq.\ (\ref{Gamma1}), color conservation,
Eq.\ (\ref{coloridentity}), implies a form in which,
as anticipated, the double poles organize themselves into
color-diagonal terms,
\bea 
\mu^{2\vep}\sum_{j\ne i}\, {\cal M}^{(1)}(p_i\cdot p_j,\ep)_{p_i^2=p_j^2=0} 
&=&  
-\
 \left( \frac{1}{2\ep}\, \left ( \sum_{j\ne i}\, {\bf T}_i\cdot {\bf
T}_j\right)\, \ln\left(\frac{-2p_i\cdot p_j}{\mu^2}\right) \ +\
\frac{1}{4\ep^2}\, C_i\right)\, . \label{eq:M1m_summed} \eea
The order $\as$ double pole terms cancel against
the form factors in the definition of the soft function. The resulting one-loop
counterterms that correspond to Eq.\ (\ref{eq:M1m_summed}) are then
pure single poles.

Going on to two loops, we follow the reasoning leading to Eq.\
(\ref{eq:Gamma-planar-ct}) in the massive case. The essential difference  is that 
all integrals remain scaleless, and we
do not encounter convergent integrals like Eq.\ (\ref{Idef}), whose
expansion in $\ep$ can give single poles at two loops.   For the
massless case, the two-loop integrals produce no expansion in $\ep$ beyond the
pole terms that define the scheme. 
Alternately, there is no pole associated with a single
scaling of all the vertices in the diagram together, of the sort we identified in 
Eq.\ (\ref{M2integrals}) with massive eikonals.   That is, in the
massless case each integral
of the two-loop diagram is fully independent.
The massless analog of Eq.\ (\ref{M2integrals})
for Fig.\ \ref{fig:cancelpair}b is then,
\bea
\mu^{4\vep}\, {\cal M}^{(2)}(p_i\cdot p_j,p_k\cdot p_j,\ep)
&=&
\mu^{4\vep}c^2(\vep)\, 
 \left ( {\bf T}_k\cdot {\bf T}_j\right)\left ( {\bf T}_i\cdot {\bf T}_j\right)\;
  \int_0^\infty d\lambda'_j \int_0^\infty d\lambda_k {p_k\cdot p_j\over
\left[-2p_k\cdot p_j \lambda_k\lambda'_j  - i \epsilon \right]^{1-\ep} }
\nonumber\\
&\ & \times
 \int_0^{\lambda_j'}
d\lambda_j \int_0^\infty d\lambda_i {p_i\cdot p_j\over
\left[-2p_i\cdot p_j \lambda_i\lambda_j  - i \epsilon \right]^{1-\ep} }\,
\bigg |_{UV}\, ,
\eea
where we assume all lightlike momenta, and where the ordering of the arguments
of ${\cal M}^{(2)}$ determines the ordering of the color matrices.
The integrand here is manifestly symmetric in $p_i$ and $p_k$.
As a result, Fig.\ \ref{fig:cancelpair}a differs from
Fig.\ \ref{fig:cancelpair}b only by the orderings of their integration
parameters $\lambda_j$ and $\lambda'_j$ 
and of their color factors.   A simple
exchange of the two integration labels eliminates the antisymmetric part in the
sum of the two diagrams, and we find
\bea
\mu^{4\vep}\, \left[\, {\cal M}^{(2)}(p_i\cdot p_j,p_k\cdot p_j,\ep)\, 
+ {\cal M}^{(2)}(p_k\cdot p_j,p_i\cdot p_j,\ep)\, \right]
=
\frac{1}{2}\, 
\left\{\, \mu^{2\vep}{\cal M}^{(1)}(p_i\cdot p_j,\ep)\,
,\, \mu^{2\vep}{\cal M}^{(1)}(p_k\cdot p_j,\ep)\, \right\}\, .
\eea
Only the symmetric part survives in the sum of the
two diagrams, considered as color matrices.   This is the coordinate space analog of a
momentum space argument leading to the same result in Ref.\ \cite{Aybat:2006mz}.
As in the massive case, we are left to analyze potentially antisymmetric contributions from
diagrams like Figs.\ \ref{fig:UVct}, with  counterterms.   
Unlike the massive case, however, the one-loop
integrals of these diagrams are scaleless in both
integration variables, and there are no finite terms,
 beyond those that
may have been included in the scheme that defines
the one-loop counterterms.   This ensures that 
the counterterm diagrams are proportional to 
the original diagrams, and that their sum remains symmetric.
In principle, these one-loop counterterms include 
double as well as single poles, but as we have seen
in our one-loop example,
double poles systematically cancel in the soft anomalous dimension matrix,
a result that extends to all orders \cite{Kidonakis:1997gm,Sen:1982bt}.

The origin of the difference between the massive and massless cases
is easy to trace to the one-loop integrals of Eqs.\ (\ref{Idef})
and (\ref{calIdef}), respectively. The integrands in the two cases differ only
when the scaled integration variable $\sigma = \lambda_i/\lambda_j$ in (\ref{Idef})
is large (or small) enough so that
\bea \frac{\lambda_i}{\lambda_j} \ge \frac{2p_i\cdot
p_j}{p_i^2} ~~~~~~ {\rm or} ~~~~~~
\frac{\lambda_j}{\lambda_i} \ge \frac{2p_i\cdot
p_j}{p_j^2} \, . \label{inequality} \eea
Only for these ``collinear" regions do the integrands distinguish between massless
and massive eikonal lines.
For any fixed mass, however, these regions are always present.
Within each diagram, therefore, the limit of zero mass does not commute with the
integrations.   This is not the case for the full soft function,
however, because these collinear regions, whether regularized by dimensions or masses,
cancel in the ratio of the full eikonal amplitude to
jet functions \cite{Kidonakis:1997gm,Mitov:2006xs,Sen:1982bt}, after the sum over 
 a gauge-invariant set of diagrams.

\section{Anomalous Dimensions Near Absolute Threshold}
\label{sec:threshold}

\begin{figure}
{\epsfxsize=4 cm \epsffile{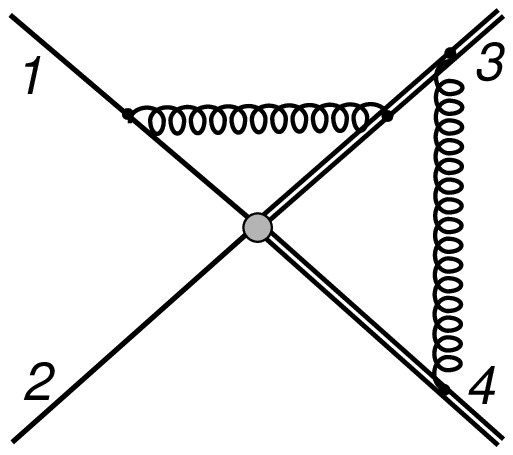} \quad \quad
\epsfxsize=4 cm \epsffile{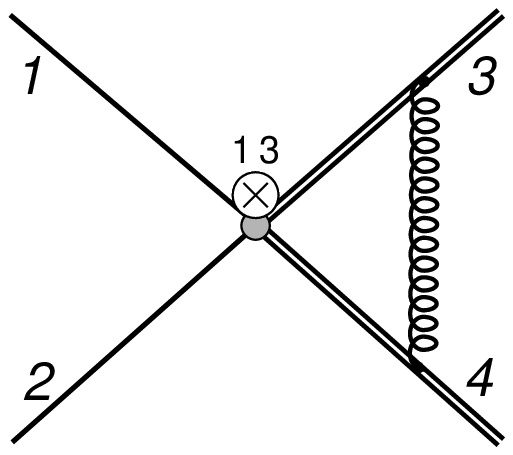} \\
\hbox{ \hskip 6.3 cm (a) \hskip 4.1  cm \quad (b) }
\vskip .1 in
\epsfxsize=4 cm \epsffile{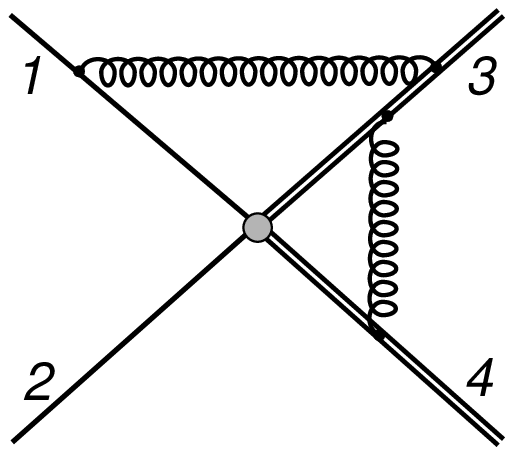} \quad \quad
\epsfxsize=4 cm \epsffile{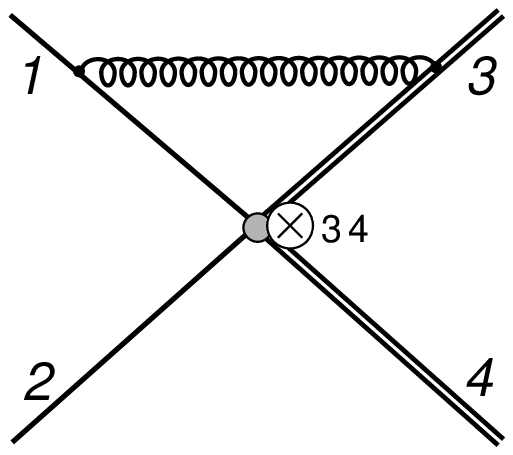}\\
\hbox{ \hskip 6.3 cm (c) \hskip 4.1  cm \quad (d) }
\caption{Two-loop double-exchange diagrams and their one-loop counterterms at order $g^4$
in the two-to-two scattering reactions $q{\bar q} \to Q{\overline
Q}$ and $gg \to Q{\overline Q}$. }
\label{fig:22}}
\end{figure}

In this section, we discuss the interpretation of threshold
behavior of the eikonal soft function representing 
a two-to-two reaction of the type $q{\bar q} \to
Q{\overline Q}$ or $gg \to Q{\overline Q}$, as in Fig.\
\ref{fig:22}. We will concentrate on the effects of the double
exchange diagrams shown there, involving (for example) an incoming
light quark or gluon, labelled 1, and the two outgoing
particles, labelled 3 and 4. Our goal here is to rederive and
interpret the non-uniform behavior in the massive anomalous dimension matrix
found in Ref.\ \cite{Ferroglia:2009ep} at absolute threshold, 
where $s\rightarrow 4M^2$, with $M$ the heavy quark mass.   We will confirm
and interpret  the conclusion \cite{Czakon:2009zw} that this behavior relates to a
region in momentum space in which the
eikonal approximation
does not apply to the underlying partonic cross section.

Equation (\ref{eq:Gamma-planar-ct}) above
shows how the double-exchange diagrams
contribute to the anomalous dimension matrix.   For 
applications to heavy quark production, as illustrated
in Fig.\ \ref{fig:22}, we take 
$p_1^2=p_2^2=m^2;~p_3^2=p_4^2=M^2$. 
We consider the incoming mass $m$ to be small, $m<<M$, and 
with heavy quark production in mind will
eventually take it to zero. 
To find the threshold, $s=(p_1+p_2)^2\rightarrow 4M^2$,
behavior for this two-to-two process, we take the 
$\beta = \sqrt{1-4M^2/s}\rightarrow 0$, with $\beta$ the center of mass velocity for the heavy pair.
\footnote{In terms of the formulas given in the appendix, this
corresponds to $x_{34}\rightarrow 1$ 
(equivalently, $v_{34}\rightarrow 0$) in Eqs.\
(\ref{eq:I-integrals-res}) and (\ref{eq:IM-parts-v}) and
$x_{13}\rightarrow 0$ (equivalently, $v_{13}\rightarrow 1$) 
 in Eqs.\ (\ref{eq:I-integrals-res-v-1}) and
(\ref{eq:I-integrals-res-v}).}   Specifically, using the expressions given in the appendix 
for $m\ll M$ we find for the terms in Eq.\ (\ref{eq:Gamma-planar-ct})
with $i=4,\, j=3,\, k=1$
corresponding to Fig.\ \ref{fig:22},
\bea 
I^{(0)}(p_4,p_3)\,
I^{(-1)}(p_1,p_3) - I^{(-1)}(p_4,p_3)\, I^{(0)}(p_1,p_3)   
&=& 
\left( 1 - \frac{i\pi}{2\beta}\right)\, \left( \ln^2\left(\frac{mM}{-t_1}\right)+\frac{\pi^2}{6}\right)
\nonumber\\
&\ & \hspace{-5mm}+\  \ln\left(\frac{mM}{-t_1}\right) 
\left[\, \frac{1}{\beta}\left( -\frac{\pi^2}{2}\; -\; i\pi\ln(4\beta)\right) +2\, \right] + {\cal O}(\beta)
\, ,
\label{eq:exchangethreshold}
\eea
where $t_1 = -2(p_1\cdot p_3)$ and $u_1 = -2(p_1\cdot p_4)$ are customary
invariants.

As noted in Sec.\  \ref{sec:massless} above, 
the double logarithm in Eq.\ (\ref{eq:exchangethreshold}), 
which originates from the collinear region, must be cancelled
by the diagram with a three-gluon interaction \cite{Ferroglia:2009ep}, 
which we reproduce in the appendix, Eq.\ (\ref{I3gspacelike}).  
After this cancellation, in the threshold limit the
diagrams of Fig.\ \ref{fig:22}  then give a contribution to the
anomalous dimension matrix, 
\bea 
\label{eq:Gamma134fig} {\bf
\Gamma}_S^{\rm (431,{\rm fig}5)} 
=
\frac{1}{2}\, if^{abc}
{\bf T}_4^a{\bf T}_3^b{\bf T}_1^c\; 
\left\{\, \ln\left( \frac{mM}{-t_1}\right)\,
\left[\, \frac{1}{\beta}\left( -\frac{\pi^2}{2}\; -\; i\pi\ln(4\beta)\right)  + 2 \,\right]  
+ \left( 1 - \frac{i\pi}{2\beta}\right)\, \frac{\pi^2}{6}\, \right\}
+ {\cal O}(\beta)\, . 
\eea
This result is still infrared divergent in the limit $m\rightarrow
0$. When we add the diagrams in which the roles of lines 3 and 4 are
reversed, however, we find the same infrared-finite
 ${\bf \Gamma}_S$ given in Eqs.~(53)-(55) of
Ref.~\cite{Ferroglia:2009ep},
\begin{equation}
\label{term-of-interest} {\bf \Gamma}_S^{\rm (431)} 
=
\frac{1}{2}\, if^{abc} {\bf T}_4^a{\bf T}_3^b{\bf T}_1^c\; \ln\left( \frac{u_1}{t_1}\right)\,
\left[\, \frac{1}{\beta}\left( -\frac{\pi^2}{2}\; -\; i\pi\ln(4\beta)\right)  + 2 \,\right]  + {\cal O}(\beta) \, . 
\end{equation}
Equation (\ref{term-of-interest}) has a very interesting behavior in
the limit $\beta \to 0$. By adopting the center of mass relation,
$t_1=-s(1-\beta\cos\theta)/2$ and similarly for $u_1$, we observe
that 
\bea
 {\bf \Gamma}_S^{\rm (431)}
 =
-\ if^{abc} {\bf T}_4^a{\bf T}_3^b{\bf T}_1^c\; 
\cos\theta\, \left( \frac{\pi^2}{2}\; +\; i\pi\ln(4\beta)\right) + {\cal O}(\beta)\  \, ,
\label{eq:fcostheta} 
\eea
in which the Coulomb singularity combines with an angular-dependent
factor that vanishes at threshold to give a term that 
remains finite for $\beta\rightarrow 0$.   As pointed out in 
Ref.\ \cite{Ferroglia:2009ep}, this term
depends on the center of mass scattering angle, no matter how small $\beta$ is.  The $\beta\rightarrow 0$
limit would thus appear to be ambiguous.  What are we to make of this?

The factorization formalism that leads to the soft anomalous dimension matrix 
\cite{Kidonakis:1997gm,Sen:1982bt} applies in the approximation
that the momentum of the active partons is much larger than that of
the soft radiation which the anomalous dimension matrix is used to
resum. The use of an anomalous dimension matrix thus requires
that the eikonal approximation apply, that is, that the emission or
absorption of this radiation  leaves the four-velocities of the
heavy quarks essentially unchanged. In fact, in the momentum region that
produces the Coulomb singularity, the center of mass kinetic energy
of the quark pair is order $m\beta^2$ in virtual as well as real
states. Soft radiation emitted by the produced pair can carry an
energy of no more than this order without violating the eikonal
approximation and washing out the $1/\beta$ dependence.   Thus, as
$\beta$ vanishes, the energy range of virtual or real radiation to
which we can apply the eikonal approximation vanishes even faster,
as the square of the relative velocity. Although the
$\beta$-independent piece of the anomalous dimension in (\ref{eq:fcostheta}), which couples singlet and
octet color states, is present  at any finite value of $\beta$, its
range of applicability shrinks to zero for $\beta\rightarrow 0$.
Correspondingly, as argued in Ref.\ \cite{Czakon:2009zw}, in the
inclusive cross section, as opposed to an elastic amplitude, it is
only the range of energies $m\ge \omega \ge m\beta^2$ that
contribute to threshold logarithms.

The Coulomb singularity at zero relative velocity $\beta$ is a
characteristic feature of heavy particle production and the analysis of
bound-state formation \cite{Hagiwara:2008df}. We can trace the origin of the $1/\beta$ 
dependence in a particularly simple fashion for the integrals
$I(p_i,p_j,\ep)$, defined in Eq.\ (\ref{Idef})  for the
exchange diagrams. Setting for definiteness $j=3,k=4$, the
integration variable $\sigma=\lambda_3/\lambda_4$ measures the
relative distances from the origin along the two heavy-particle
eikonal lines, with momenta $p_3$ and $p_4$. The
$\sigma$ integration contour encounters two singularities, at
\bea \sigma_{\pm} &=& \frac{ p_3\cdot p_4}{m^2} \pm
\frac{1}{m^2}\sqrt{(p_3\cdot p_4)^2-m^4} \pm i\epsilon
\nonumber\\
&=& 1 \pm 2\beta  \pm i\epsilon\ + {\cal O}\left (\beta^2\right)\, . \eea
In the limit of vanishing relative velocity, then, the $\sigma$
integral is pinched between coalescing singularities at $\sigma=1$,
whose separation vanishes linearly with $\beta$.  In terms of the original integrals over
$\lambda_3$ and $\lambda_4$, the singularity at $\sigma=1\pm 2\beta$
corresponds to singularities in (say) $\lambda_3$ at
$\lambda_3=\lambda_4(1 \pm 2\beta)$. 
In particular, for $\beta=0$, the two ends of the gluon propagator
are at the same point in coordinate space, $\lambda_3p_3=\lambda_4p_4$.
We can use this analysis to interpret further
the range of applicability of the anomalous dimension
matrix.

Working in the $p_3,\ p_4$ center of mass frame, the quantities
$\lambda_3+\lambda_4$ and $\lambda_3-\lambda_4$ are proportional to
the temporal and spatial distances between two points along the $p_3$
and $p_4$ eikonals, and are conjugate to the energy and spatial
momentum of emitted radiation. In the region that dominates the
integrals, the conjugate of the spatial momenta is constrained to be
smaller than the conjugate of the energy by a factor of $\beta$,
corresponding to a range in spatial momentum larger than energy by a
factor of $1/\beta$. Thus, when the energy of gluon exchange
approaches the scale of heavy quark kinetic energy, $m\beta^2$, the
exchanged momentum approaches $m\beta$, which is the same order as
the heavy quark momentum. In this region, the quarks can no
longer by considered as recoilless sources, and the eikonal
approximation is not reliable.  In Ref.\ \cite{Beneke:2009rj}, an
effective theory treatment for resummation in this region was
developed, and in \cite{Czakon:2009zw} an equivalent analysis in
which the pair is replaced by a single Wilson line was employed. In
either case, it is necessary to assume that radiation with
energy-momentum scales up to $m\beta$ cancels in the sum over final
states for the inclusive cross section.

\begin{figure}
{\epsfxsize=4.5 cm \epsffile{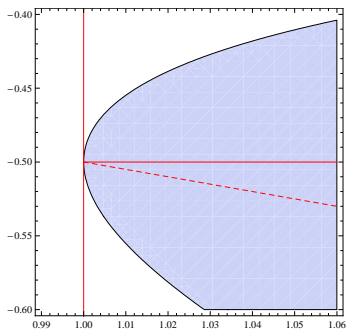} \\
\caption{Illustration of the physical region (shaded) in the
$(s,t_1)$ kinematic plane. The normalization is chosen such that
$s_0=1$. The solid lines depict the lines $\beta=0$ (i.e. $s=s_0$)
and $t=t_0$. The dashed line denotes the set of points $t_1=-s/2$
where the function ${\bf \Gamma}_S^{\rm (134)}$, defined in
Eq.~(\ref{term-of-interest}), vanishes due to antisymmetry (note
that this line also crosses the absolute threshold point
$t_0=-s_0/2$).} \label{fig:s-t-plane}}
\end{figure}

Finally, it is interesting to note that the connection between
$\cos\theta$ and the kinematic variables $s$ and $t_1$ used in the
derivation of Eq.~(\ref{eq:fcostheta}) is singular at absolute
threshold. In two-to-two scattering processes, $s+t_1+u_1=0$, and
the function ${\bf \Gamma}_S^{\rm (134)}$ in
Eq.~(\ref{term-of-interest}) is determined by two variables, which
we can choose as ($s,t_1$). The absolute threshold $\beta=0$ then
represents the single point $(s=s_0\equiv 4m^2, t_1=t_0\equiv
-s_0/2)$ in the physical region of the kinematical $(s,t_1)$ plane;
see Fig.~\ref{fig:s-t-plane}.
The relationship between $t_1$ and $s$ can now be written as
\bea t_1 &=& -\frac{s}{2}(1-\beta\cos\theta) \ =\
-\frac{s_0}{2}+\frac{\sqrt{s}}{2}\, \sqrt{s-s_0}\, \cos\theta -
\frac{1}{2}\left(s-s_0\right)\, . 
\eea
At threshold, all derivatives of $t_1$ with respect to $s$ diverge
unless $\cos\theta=0$. Therefore, in terms of figure
\ref{fig:s-t-plane}, all curves that approach absolute threshold at
fixed, non-zero $\cos\theta$ converge to the vertical
tangent to the boundary of the physical region at the point
$t_1=t_0,\, s=s_0$. At the same time, it is straightforward to
verify that along  any straight line defined by the equation
\bea
t_1 = t_0 +c\left(s-s_0\right) \, , \quad u_1= t_0 - (1+c)(s-s_0)
\, , \quad c < \infty \,\, ,
\eea
the function ${\bf \Gamma}_S^{\rm (134)}$ in
Eq.~(\ref{term-of-interest}) vanishes at threshold and thus matches
smoothly to the behavior of the function at $u_1=t_1$
\cite{Czakon:2009zw}. In a sense, as $\beta\rightarrow 0$, the
cosine becomes ill-defined,
because it no longer uniquely defines the Mandelstam variables. At
the same time, for any fixed value of $\beta$, and in particular,
for $\beta\rightarrow 1$, the full expression for the soft anomalous
dimension as found in Ref.\ \cite{Ferroglia:2009ep} applies without
subtlety for gluon radiation in the range identified above.

\section{Conclusions}

In this note we have demonstrated the mutual consistency of a full
position-space calculation of the massive soft anomalous
dimension matrix following Ref.~\cite{Mitov:2009sv},
and the momentum space results described in
Ref.~\cite{Ferroglia:2009ep}. In particular, 
the color-symmetric structure of the two-loop
double-exchange diagrams  before renormalization 
shown in \cite{Mitov:2009sv}
is consistent with  the results for planar
diagrams found in Ref.~\cite{Ferroglia:2009ep}, once renormalization
is taken into account.   
We have also confirmed the consistency of results
for the massive anomalous dimension matrix with those for the massless
case and with the next-to-next-to leading logarithmic threshold
resummations described in  \cite{Beneke:2009rj,Czakon:2009zw}.
Finally, we have seen how coordinate space analysis
provides a clear interpretation of the non-uniform
limit of the anomalous dimension matrix at absolute threshold.

\acknowledgments A.M. would like to thank Michal~Czakon, Kirill~Melnikov and
Shlomo~Razamat for  discussions. This work was
supported in part by the National Science Foundation, grants
PHY-0354776, PHY-0354822 and PHY-0653342. The work of A.M. is supported by a
fellowship from the {\it US LHC Theory Initiative} through NSF grant
0705682.

\appendix\section{Explicit forms for the
integrals}\label{sec:appendix}

The functions $I(p_i,p_j,\ep)$, defined by Eq.\ (\ref{Idef}), are
one-dimensional integrals. For space-like kinematics (for example
$p_i$ incoming, $p_j$ outgoing) the result in $d=4-2\ep$ dimensions
and for arbitrary masses
can be expressed as
\begin{equation}
I(p_i,p_j,\ep) = {1\over 1-2\ep} ~ {_2F_1}\left( {1\over 2},1;
{3\over 2}-\ep ; 1-{m_i^2m_j^2\over (p_i\cdot p_j)^2}\right) \, .
\label{eq:integral-2F1}
\end{equation}
The above function has a well defined massless limit
in $4-2\ep$ dimensions: ${_2F_1}\left( 1/2,1; 3/2-\ep ; 1\right) =
-(1-2\ep)/(2\ep)$. As can be seen from Eq.~(\ref{eq:integral-2F1})
all non-trivial dependence on the kinematical invariants and masses
comes entirely through the combination:
\begin{equation}
\vij = \sqrt{1-{m_i^2 m_j^2\over (p_i\cdot p_j)^2}} \, ,
\end{equation}
familiar from the one-loop case \cite{Catani:2000ef}. Expressions
for the functions $I^{(-1,0)}$ appearing in
Eqs.~(\ref{eq:Gamma-planar-ct}), (\ref{eq:Gamma-full-ct}) can be
obtained
by
expanding the hypergeometric function in $\ep$, or by
expansion of the integrand in Eq.\ (\ref{Idef}) before integration.
For expansion of the full expression (\ref{eq:integral-2F1}) we have
used the program {\it HypExp} \cite{Huber:2005yg}, as well as
standard relations between the polylogarithmic functions. The
resulting expressions can be written as
\begin{eqnarray}
I^{(-1)}(p_i,p_j) &=& - {1\over 2\vij}~\ln\left({1-\vij\over
1+\vij}\right)\, ,
\label{eq:I-integrals-res-v-1}\\
I^{(0)}(p_i,p_j) &=&  {1\over \vij}~\Bigg\{ -{\rm
Li}_2\left({1-\vij\over 1+\vij}\right)  + {1\over
4}\ln^2\left({1-\vij\over 1+\vij}\right) +\ln\left({1-\vij\over
1+\vij}\right)\ln\left({1+\vij\over 2\vij}\right) +{\pi^2 \over 6}
\Bigg\}\, . \label{eq:I-integrals-res-v}
\end{eqnarray}
The consistency of these results with those of Ref.\
\cite{Ferroglia:2009ep} is readily checked by relating the variable
$\beta_{ij}=-\cosh^{-1}(\pm p_i\cdot p_j/m_im_j)$ to our $v_{ij}$ by
\bea \beta_{ij}= - \frac{1}{2}\,
\ln\left(\frac{1-v_{ij}}{1+v_{ij}}\right)\, , \quad \coth\beta_{ij}
= \frac{1}{v_{ij}}\, . \eea

While compact, the expressions in Eqs.\
(\ref{eq:I-integrals-res-v-1}) and (\ref{eq:I-integrals-res-v}) are somewhat
inconvenient for analytic continuation from space-like to time-like
kinematics, because their momentum-dependence is only through the
squares $(p_i\cdot p_j)^2$ in $v_{ij}$. For completeness and for use
in our discussion of the threshold limit, we recall that analytic
continuation is made straightforward by re-expressing the amplitudes
in terms of the variable
\bea \label{eq:x-deff} \xij &=& \sqrt{\frac{1-v_{ij}}{1+v_{ij}}}
\ =\ { \sqrt{1-{(m_i+m_j)^2\over s_{ij}}} -
\sqrt{1-{(m_i-m_j)^2\over s_{ij}}} \over \sqrt{1-{(m_i+m_j)^2\over
s_{ij}}} + \sqrt{1-{(m_i-m_j)^2\over s_{ij}}}} \,\, ,
 \quad v_{ij}= {1-\xij^2\over 1+\xij^2}\, ,
\eea
with $s_{ij}=(p_i+p_j)^2$.  The case of $m_i=m_j=m$ is of particular
interest, where
\begin{equation}
\label{eq:x-deffequal} \xij = { \sqrt{1-{4m^2\over s_{ij}}} - 1
\over \sqrt{1-{4m^2\over s_{ij}}} + 1} = -\
\frac{1-b_{ij}}{1+b_{ij}}\,\, , \quad \quad  {m^2\over s_{ij}} = -
{\xij\over (1-\xij)^2}\, ,
\end{equation}
where $b_{ij}$ is the center-of-mass velocity for the pair $p_i$,
$p_j$.    For arbitrary masses, we have
$x_{ij}>0$ for $s_{ij}<0$ (space-like) and
$x_{ij}<0$ for $s_{ij}>0$ (time-like).

For any masses, we now rewrite the functions
$I^{(-1,0)}$ in terms of the $x_{ij}$ as
\begin{eqnarray}
I^{(-1)}(p_i,p_j) &=& - {1+\xij^2\over 1-\xij^2}~\ln(\xij)\,
,\label{eq:I-integrals-res-1}\\
I^{(0)}(p_i,p_j) &=&  {1+\xij^2\over 1-\xij^2}~\left( - {\rm
Li}_2(\xij^2) + \ln^2(\xij) - 2\ln(\xij)\ln(1-\xij^2) + {\pi^2\over
6} \right)\, . \label{eq:I-integrals-res}
\end{eqnarray}
For space-like kinematics, where $1>x_{ij}>0$, these functions are real. Analytic
continuation to time-like kinematics is found by the replacement:
$\xij \to -|\xij| +i\varepsilon = |\xij|e^{i\pi}$ (see Sec.\ 6 of Ref.\ \cite{Bernreuther:2004ih}
for details).  Following this rule, the following terms should be added to
Eqs.~(\ref{eq:I-integrals-res-v-1}), (\ref{eq:I-integrals-res-v})
when the kinematics is time-like:
\begin{eqnarray}
\Delta I^{(-1)}(p_i,p_j) &=& - i\pi\, {1+\xij^2\over 1-\xij^2}\ =\
-\ {i\pi\over \vij}\, ,
\label{eq:IM-parts-v-1}\\
\Delta I^{(0)}(p_i,p_j) &=&  {1+\xij^2\over 1-\xij^2}\, \left(\, -
\pi^2  + i\pi\, \left[ 2\ln{|\xij|} - 2\ln\left({1-\xij^2}\right)
\right]\, \right)
\nonumber\\
&=&   -\ {\pi^2\over \vij}
 + {i\pi\over \vij}\left[ \ln\left({1-\vij\over 1+\vij}\right) +
2\ln\left({1+\vij\over 2\vij}\right) \right] \, .
\label{eq:IM-parts-v}
\end{eqnarray}

Finally, for space-like kinematics the result for the non-planar
diagram in Fig.~\ref{nonplanar},
as expressed in Eq.\ (\ref{eq:Gamma-full-ct}) above, reads \cite{Ferroglia:2009ep}:
\begin{equation}
I_{3g}(p_i,p_j) = -\ln^2(\xij)  = -{1\over
4}\ln^2\left({1-\vij\over 1+\vij}\right)  \, .
\label{I3gspacelike}
\end{equation}
In time-like kinematics, it also receives the following term:
\begin{equation}
\Delta I_{3g}(p_i,p_j) = \pi^2 - 2i\pi\ln(|\xij|) = \pi^2 -
i\pi\ln\left({1-\vij\over 1+\vij}\right) \, .
\end{equation}

\end{document}